\documentstyle[12pt]{article}
\begin{document}
\begin{titlepage}
\mbox{ }
\rightline{UCT-TP-219/94}
\rightline{October 1994}
\vspace{1.5cm}
\begin{center}
\begin{Large}
{\bf  Heavy quark masses from QCD sum rules\footnote{Invited
talk at the Mexican School of Particles and Fields, Villahermosa,
Mexico, October 1994.}}

\end{Large}

\vspace{2cm}

{\large {\bf C. A. Dominguez \footnote{John Simon Guggenheim Fellow
1994-1995.} }}

Institute of Theoretical Physics and Astrophysics, University of
Cape Town, Rondebosch 7700, South Africa\\
\end{center}

\vspace{1.2cm}

\begin{abstract}
\noindent
I review recent determinations of the (on-shell) charm- and
beauty-quark masses in the framework of relativistic and non-relativistic
ratios of Laplace transform QCD moment sum rules. The validity of the
non-relativistic version of QCD sum rules in this particular application
is discussed.
\end{abstract}
\end{titlepage}

\setlength{\baselineskip}{1\baselineskip}
\noindent
With the advent of the Heavy Quark Effective Theory (HQET) there has been a
revived interest in applications of QCD sum rules to the heavy quark
sector. The purpose being the update of old determinations, as well as
the performance of new calculations to extract accurate values of
various dynamical quantities entering the HQET. Chief of
these quantities is the ubiquitous  heavy-quark mass.
I review here recent determinations of the (on-shell) charm- and
beauty-quark masses  \cite{1} performed by confronting
very accurate experimental data  on the charmonium and the
upsilon systems \cite{2} with ratios of  relativistic and non-relativistic
Laplace transform QCD moments. The latter theoretical framework, suggested
by Bertlmann \cite{3}, offers several advantages, e.g. radiative
and non-perturbative corrections are well under control, and the
non-relativistic limit follows quite naturally  from quantum mechanical
analogues \cite{4}. This version of QCD sum rules leads to an expansion
in powers of the inverse of the heavy quark mass which allows one to test
the range of validity of the non-relativistic limit, and more generally, to
assess the role of  mass corrections. This might be of interest for
calculations based on the simplifying assumption $\Lambda_{QCD}/m_Q\ll 1$.
Also, a comparison of the results from the relativistic and the
non-relativistic determinations provides an estimate of the systematic
uncertainties affecting this technique.

I begin by considering the two-point function
\begin{equation}
\Pi_{\mu \nu} (q) =  i \; \int d^{4}x \exp (i q x) \langle0|T(V_{\mu} (x)
V^{+}_{\nu}
(0))|0 \rangle
		   =  (- g_{\mu \nu} q^{2} + q_{\mu} q_{\nu}) \Pi
		  (q^{2}) \; ,
\end{equation}
with $V_{\mu}(x)=\bar{Q}(x) \gamma_{\mu} Q(x)$. The function $\Pi(q^{2})$
has been calculated in perturbative QCD at the two-loop level \cite{5},
with its imaginary part given by
\begin{equation}
\frac{1}{\pi} \; \mbox{Im} \; \Pi (s)|_{QCD} = \frac{1}{8 \pi^{2}}
v(3 - v^{2})
\Biggl\{ 1 + \frac{4 \alpha_{s}}{3} \left[ \frac{\pi}{2v} - \frac{(v+3)}{4}
\left( \frac{\pi}{2} - \frac{3}{4 \pi} \right) \right] \Biggr\}
\theta (s - 4 m_{Q}^{2}) \; ,
\end{equation}
where $v=\sqrt{1 - 4 m_{Q}^{2}/s}$, and $m_{Q}$ is the charm- or the
beauty-quark on-shell mass:  $m_{Q}=m_{Q}(Q^{2}=m_{Q}^{2})$.
The leading non-perturbative term in the operator product expansion
of   $\Pi(q^{2})$ involves the gluon condensate, i.e.
\begin{equation}
\Pi (s)|_{NP} = \frac{1}{48 \; s^{2}} \times \left[
\frac{3(v^{2} + 1)(1 - v^{2})^{2}}{2 v^{5}} \ln \frac{1+v}{1-v}
- \frac{3v^{4} - 2v^{2} + 3}{v^{4}} \right]
\langle \frac{\alpha_{s}}{\pi} G^{2} \rangle \; .
\end{equation}
The function $\Pi(q^{2})$ satisfies a once-subtracted dispersion relation,
and the subtraction constant can be disposed of e.g. by taking the Laplace
transform
\begin{equation}
\Pi (\sigma) =  \int_{0}^{\infty} ds \exp (- \sigma s)
\; \mbox{Im} \; \Pi (s)   \; .
\end{equation}
The quantity of interest to us here is the ratio of the first two Laplace
moments, which can be expressed as
\begin{equation}
{\cal{R}} (\sigma) = - \frac{d}{d \sigma} \ln \; \Pi (\sigma) \; .
\end{equation}
Substituting Eq.(2) into Eq.(4) one can carry out the integration
analytically. The result for the ratio Eq.(5), with
$\omega = 4 m_{Q}^{2} \sigma$, is \cite{3}
\begin{equation}
{\cal{R}} (\omega) = 4 m_{Q}^{2} \left[ 1 - \frac{A'(\omega)}{A(\omega)} -
\frac{a'(\omega) \alpha_{s} + b'(\omega) \phi}
{1 + a(\omega) \alpha_{s} + b (\omega) \phi} \right] \; ,
\end{equation}
where
\begin{equation}
\pi A(\omega) = \frac{3}{16 \sqrt{\pi}} \frac {4 m_{Q}^{2}}{\omega}
\; G(\frac{1}{2}, \frac{5}{2}, \omega) \; ,
\end{equation}
\begin{equation}
a(\omega) = \frac{4}{3\sqrt{\pi}} \; G^{-1} (\frac{1}{2}, \frac{5}{2},
\omega)
[\pi - c_{1} \; G(1,2, \omega) + \frac{1}{3} c_{2} \; G(2,3,\omega)]
- c_{2} \; ,
\end{equation}
\begin{equation}
b(\omega) = - \frac{\omega^{2}}{2} \; G(-\frac{1}{2}, \frac{3}{2}, \omega)
\; G^{-1}(\frac{1}{2}, \frac{5}{2}, \omega) \; ,
\end{equation}
\begin{equation}
c_{1} = \frac{\pi}{3} + \frac{c_{2}}{2} \; \; \; , \; \; \;
c_{2} = \frac{\pi}{2} - \frac{3}{4 \pi} \; ,
\end{equation}
\begin{equation}
A'(\omega) = - \frac{A(\omega)}{\omega} \left[ \frac{3}{2} -
\frac{5}{4} G(\frac{3}{2}, \frac{7}{2}, \omega) \; G^{-1} (\frac{1}{2},
\frac{5}{2}, \omega) \right] \; ,
\end{equation}
\begin{eqnarray*}
a'(\omega) = \frac{4}{3 \omega \sqrt{\pi}} \; G^{-1} (\frac{1}{2},
\frac{5}{2}, \omega) \Biggl\{ \frac{1}{2} \; G^{-1} (\frac{1}{2},
\frac{5}{2},
\omega) \; \left[ G(\frac{1}{2}, \frac{5}{2}, \omega)  \right. \Biggr.
\end{eqnarray*}
\begin{eqnarray*}
\left. - \frac{5}{2} G( \frac{3}{2}, \frac{7}{2}, \omega)
\right] \left[ \pi - c_{1} G(1,2, \omega) + \frac{c_{2}}{3}
G(2,3, \omega) \right]
\end{eqnarray*}
\begin{equation}
\left. + c_{1} [G(1,2, \omega) - 2 G(2,3, \omega)] + \frac{1}{3} c_{2}
[- 2 G(2,3, \omega) + 6 G(3,4, \omega)] \right\} \; ,
\end{equation}
\begin{eqnarray*}
b'(\omega) = \frac{2}{\omega} b(\omega) - \frac{\omega}{4}
\left\{ G(- \frac{1}{2}, \frac{3}{2}, \omega)
G^{-1} (\frac{1}{2}, \frac{5}{2}, \omega) \right.
\end{eqnarray*}
\begin{equation}
\left.  - \frac{3}{2} + G(- \frac{1}{2}, \frac{3}{2}, \omega)
G^{-2} (\frac{1}{2}, \frac{5}{2}, \omega)
\left[ G(\frac{1}{2}, \frac{5}{2}, \omega) - \frac{5}{2}
G(\frac{3}{2}, \frac{7}{2}, \omega)  \right] \right\} \; ,
\end{equation}
\begin{equation}
\alpha_{s} (Q^{2}) = \frac{-2 \pi}{\beta_{1} \;
\ln \; Q^{2}/\Lambda^{2}} \; ,
\end{equation}
\begin{equation}
\phi = \frac{\pi}{36} \frac{<\alpha_{s} G^{2}>}{m_{Q}^{4}} \; ,
\end{equation}
\begin{equation}
G(b,c,\omega) = \frac{\omega^{-b}}{\Gamma (c)} \int_{0}^{\infty}
dt \; t^{c-1} e^{-t} (1 + \frac{t}{\omega})^{-b} \; .
\end{equation}
The function $G(b,c,\omega)$ is related to the Whittaker function
$W_{\lambda,\mu}(\omega)$ through \cite{6}
\begin{equation}
G(b,c,\omega) = \omega^{\mu - 1/2} e^{\omega/2} \; W_{\lambda,\mu}
(\omega) \; ,
\end{equation}
with $\mu=(c-b)/2$, and $\lambda =(1-c-b)/2$.

The above expressions involve no approximations, other than the
two-loop perturbative expansion, and the truncation of the operator
product expansion beyond  the leading non-perturbative term.
We shall refer to Eq.(6) as the fully relativistic Laplace ratio.
In the non-relativistic (heavy quark-mass) limit, the Laplace transform
Eq.(4) becomes
\begin{equation}
\Pi(\tau) =  \int_{0}^{\infty} dE \exp (- \tau E)
\; \mbox{Im} \;
\Pi(E) \; ,
\end{equation}
where $\tau = 4 m_{Q} \sigma$, and $s = (2 m_{Q} + E)^{2}$ so that
$E \geq 0$. The Laplace ratio Eq.(5) is now given by
\begin{equation}
{\cal{R}}(\tau) = 2 m_{Q} - \frac{d}{d \tau} \; \ln \; \Pi(\tau) \; .
\end{equation}
After expanding the functions $G(b,c,\omega)$ entering Eq.(6),
one obtains the non-relativistic ratio
\begin{eqnarray*}
{\cal{R}}(\tau) = 2 m_{Q} \left\{ 1 + \frac{3}{4} \frac{1}{m_{Q} \tau}
\left( 1 - \frac{5}{6} \frac{1}{m_{Q} \tau} + \frac{10}{3}
\frac{1}{m_{Q}^{2} \tau^{2}} \right) \right.
\end{eqnarray*}
\begin{eqnarray*}
- \frac{\sqrt{\pi}}{3} \frac{\alpha_{s}}{\sqrt{m_{Q} \tau}}
\left[ 1 - \left( \frac{2}{3} + \frac{3}{8 \pi^{2}} \right)
\frac{1}{m_{Q} \tau} + \frac{1}{32} \left(107 + \frac{51}{\pi^{2}}
\right) \frac{1}{m_{Q}^{2} \tau^{2}} \right]
\end{eqnarray*}
\begin{equation}
\left.
+ \frac{\pi}{48} \frac{\tau^{2}}{m_{Q}^{2}} \langle \alpha_{s} G^{2}
\rangle \left( 1 + \frac{4}{3} \frac{1}{m_{Q} \tau} - \frac{5}{12}
\frac{1}{m_{Q}^{2} \tau^{2}} \right) \right\} \; ,
\end{equation}
where the expansion has been truncated at the next-to-next to leading
order in $1/m_{Q}$. Notice that the appearance of $\sqrt{m_{Q}}$
above is only an artifact of the change of variables; written
in terms of $\sigma$, Eq.(20) contains no such term.
The theoretical ratios of the first two Laplace moments Eqs.(6) and (20)
must now be confronted with a corresponding ratio involving the experimental
data on the $J/\psi$  and the $\Upsilon$ systems.
In the case of the $J/\psi$  one parametrizes the data by a sum of
two narrow resonances below $D\bar{D}$ threshold, followed by
a hadronic continuum modelled by perturbative QCD. In the case of
the $\Upsilon$, three narrow resonances below the $B\bar{B}$ threshold
are required, at least in principle. One obtains
\begin{equation}
\Pi (\sigma)|_{EXP} = \frac{3}{4 \pi} \frac{1}{e_{Q}^{2} \alpha_{EM}^{2}}
\sum_{V} \Gamma_{V}^{ee} M_{V} \exp (- \sigma M_{V}^{2})
+ \frac{1}{\pi} \int_{s_{0}}^{\infty} ds \; \exp (- \sigma s) \; \mbox{Im}
\; \Pi (s)|_{QCD} \; .
\end{equation}
The experimental ratio is then calculated using Eq.(21) in Eq.(5). The
continuum threshold $s_{0}$ is chosen at or below the $D\bar{D}$
($B\bar{B}$) threshold. Reasonable changes in the value
of $s_{0}$ have essentially no impact on the results, as $\Pi(\sigma)$
is saturated almost entirely by the first two $J/\psi$ narrow resonances
in the case of charm, and the first $\Upsilon$ state in the case of beauty.
In the theoretical ratios the following current values
of the QCD parameters have been used:
$\Lambda = 200 - 300 \; \mbox{MeV}$, for four flavours, and
$\Lambda = 100 - 200\; \mbox{MeV}$, for five flavours \cite{2}, and
$<\alpha_{s} G^{2}> = 0.063 - 0.19 \;\mbox{GeV}^{4}$  \cite{7}.

In the case of charm, one finds that theoretical and experimental
ratios match in the wide sum rule window: $\sigma \simeq 0.8 - 1.5 \;$
$\;\mbox{GeV}^{- 2}$, for $m_{c} = 1.39 - 1.46 \;\mbox{GeV}$ in the fully
relativistic case, and  $\sigma \simeq 0.6 - 0.8 \;\mbox{GeV}^{-2}$,
$m_{c} = 1.40 - 1.53 \;\mbox{GeV}$ in the non-relativistic case.
For values of $\sigma$ inside the sum rule window, the hierarchy of the
various terms in the non-relativistic Laplace ratio (20) guarantees a
fast convergence. In fact, the leading correction in $1/m_{c}$ is at
the 15-20\% level, the radiative correction and the non-perturbative
contribution amount both to less than 10\% . At the same time,
the next, and next-to-next to leading (in $1/m_{c}$) terms everywhere in
Eq.(20) are safely small, as it can be easily verified from Eq.(20) noticing
that if $\sigma \simeq 1/2 \;\mbox{GeV}^{-2}$, then $\tau \simeq 2 m_{c}$.
Clearly, the complete analysis at the level of accuracy of these
next-to-leading mass corrections would require the
evaluation of the perturbative $O(\alpha_s^2)$ terms.
Combining the results from both versions of the Laplace ratios, leads to
the result
\begin{equation}
m_{c} (Q^{2} = m_{c}^{2}) = 1.46 \pm 0.07 \; \mbox{\mbox{\mbox{GeV}}}\; .
\end{equation}
In the case of beauty, at small and intermediate values of $\sigma$ the
$\Upsilon (1S)$ provides the bulk of the hadronic contribution, i.e.
the $\Upsilon (2S)$, $\Upsilon (3S)$, and the continuum represent a
small correction, below the spread in the theoretical ratio due to
variations in $\Lambda$ and in $<\alpha_{s}G^{2}>$. Theoretical and
experimental ratios match inside the wide regions:
$\sigma \simeq 0.4 - 0.8 \;\mbox{GeV}^{-2}$, for $m_{b} = 4.63 - 4.67\;
\mbox{GeV}$ in the fully relativistic case,
and $\sigma \simeq 0.20 - 0.35 \;\mbox{GeV}^{-2}$, for $m_{b} = 4.69 - 4.77
\;\mbox{GeV}$ in the non-relativistic one. In the latter case all
correction terms in Eq.(20) are at the safe level of a few percent. The
subleading quark mass corrections, though small, are important. For
instance, the term of order ${\cal O} (\alpha_{s}/m_{b} \sqrt{m_{b}})$
in Eq.(20) is of the same size and sign as the non-perturbative term.
Hence, it is not fully justified to keep the latter and ignore the former.
After combining the results from both methods one predicts
\begin{equation}
m_{b} (Q^{2} = m_{b}^{2}) = 4.70 \pm 0.07 \; \mbox{\mbox{\mbox{GeV}}}\; .
\end{equation}
When comparing the results reported here, Eqs.(22)-(23), with previous
determinations based on various versions of QCD sum rules \cite{3},
\cite{8} - \cite{9}, it is important to know which values of $\Lambda$
and $<\alpha_{s} G^{2}>$ have been used, as well as which renormalization
point has been chosen, e.g. some authors determine $m_{Q}(Q^{2}=
- m_{Q}^{2})$,  which is related to the on-shell mass
$m_{Q}(m_{Q}^{2})$ through
\begin{equation}
m_{Q}^{2}(m_{Q}^{2}) = m_{Q}^{2}(-m_{Q}^{2})
(1 + \frac{4 \; \ln \; 2}{\pi} \alpha_{s}) \; .
\end{equation}
The results from the present method are in very good agreement with those
of \cite{3} and \cite{8}. Comparison with other analyses \cite{9} is
often made difficult by the lack of information on the specific
values used for $\Lambda$ and the gluon condensate.\\
{\bf Acknowledgements}\\
The author wishes to thank the organizers of the Mexican School
of Particles and Fields for a very successful
meeting. This work was supported in part by the Foundation for Research
Development (ZA) and the John Simon Guggenheim Memorial Foundation (USA).

\end{document}